\documentclass[aps, reprint,longbibliography]{revtex4-1}

\usepackage{graphicx}
\usepackage{hyperref}
\hypersetup{
    bookmarks=false,         
    unicode=false,          
    pdftoolbar=false,        
    pdfmenubar=true,        
    pdffitwindow=false,     
    pdfstartview={FitH},    
    pdftitle={},    
    pdfauthor={},     
    pdfsubject={},   
    pdfcreator={},   
    pdfproducer={}, 
    pdfkeywords={sudden approximation} {molecular rotations}, 
    pdfnewwindow=true,      
    colorlinks=true,       
    linkcolor=black,          
    citecolor=magenta,        
    filecolor=magenta,      
    urlcolor=magenta
}

\usepackage[normalem]{ulem}
\PassOptionsToPackage{usenames,dvipsnames}{xcolor}
\usepackage{color}
\usepackage{placeins}
\usepackage{bm}
\usepackage{amsmath, amsthm, amssymb}
\usepackage{mathtools, xfrac}

\renewcommand{\and}{\quad \text{and}\quad}
\newcommand{\ie}{i.\,e. }

\renewcommand{\d}{\mathrm{d}}

\newcommand{\im}{\mathrm{i}\mkern1mu}

\begin{document}
\title{Modeling laser pulses as $\delta$-kicks: reevaluating the impulsive limit\\ in molecular rotational dynamics}
\author{Volker Karle}
\email{volker.karle@ist.ac.at}
\author{Mikhail Lemeshko}
\email{mikhail.lemeshko@ist.ac.at}
\affiliation{Institute of Science and Technology Austria, Am Campus 1, 3400 Klosterneuburg, Austria}
\date{July 14, 2023}
\begin{abstract}
The impulsive limit (the ``sudden approximation'') has been widely employed to describe the interaction between molecules and short, far-off-resonant laser pulses. This approximation assumes that the timescale of the laser--molecule interaction is significantly shorter than the internal rotational period of the molecule, resulting in the rotational motion being instantaneously ``frozen'' during the interaction. This simplified description  of  laser--molecule interaction is incorporated in
various theoretical models predicting  rotational dynamics of molecules driven by short laser pulses.  In this theoretical work, we develop an effective theory for ultrashort laser pulses by examining the full time-evolution operator and solving the time-dependent Schrödinger equation at the operator level. Our findings reveal a critical angular momentum,  $l_\mathrm{crit}$, at which the impulsive limit breaks down. In other words, the validity of the sudden approximation depends not only on the
pulse duration, but also on its intensity, since the latter determines how many angular momentum states are populated. We explore both ultrashort multi-cycle (Gaussian) pulses and the somewhat less studied half-cycle pulses, which produce distinct effective potentials. We discuss the limitations of the impulsive limit and propose a new method that rescales the effective matrix elements, enabling an improved and more accurate description of laser--molecule interactions.  
\end{abstract}
\maketitle
\section{Introduction}
The control and manipulation of molecules with laser pulses is of paramount importance in diverse fields such as spectroscopy, chemistry, materials science, quantum optics, and even biology~\cite{KochRMP19}. A comprehensive understanding of the post-pulse rotational dynamics of molecules is vital for the development of new technologies, including ultrafast spectroscopy and laser-induced chemistry~\cite{lin2018all,gershnabel2010controlling,milner2016laser}. Additionally, the rotational degrees of freedom of a molecule have the potential to serve as a new platform for qubits, the fundamental building blocks of quantum computing and quantum memory~\cite{KarraJCP16, YuNJP19}.\\
\\
Since the Born-Oppenheimer approximation (separation of the electronic, vibrational, and rotational  timescales) works well for many of the small molecules, at low energies they can be reliably described as quantized rigid rotors~\cite{LevebvreBrionField2}. For off-resonant ultrashort laser pulses (usually with infrared frequencies far detuned from any transitions), the rotational motion is generally considered to be slow compared to the laser modes, leading to the ``frozen'' rotational motion assumption during the laser--molecule interaction~\cite{Ortigoso1999,CaiPRL01,cai2001recurring, dion2001orienting,leibscher2003molecular,fleischer2009controlling,seideman1999revival,stapelfeldt2003colloquium}. This justifies the impulsive limit, which adapts a semi-classical approach by neglecting the accumulation of quantum phases during the pulse duration.\\
\\
For quantum rotors, however, the energy splittings grow linearly with the angular momentum $l$, which causes the corresponding change of the relevant timescales. Therefore the applicability of the sudden approximation does not solely rely on the duration of the laser pulse, but also on its intensity which determines how many $l$-states are populated during the laser excitation. For example, for a molecule with a rotational period $\tau_\mathrm{rot}(l)$,  for the impulsive limit to be valid, only states with $l$ satisfying $\tau_\mathrm{rot}(l) \gg \tau_L$ should be occupied, where $\tau_L$ represents the pulse duration of the laser. Additionally, the specific shape of the laser pulse is an important factor to consider. It is not immediately evident which values of $\tau_\mathrm{rot}(l)$ are large enough or how different laser shapes affect this relationship. Despite the widespread adoption of the impulsive limit as a theoretical framework to describe molecular rotational response to a laser pulse, a comprehensive analysis of the specific states for which this approximation is valid remains unexplored. \\\\
In this work, we aim to develop an effective theory for ultrashort laser pulses by analyzing the full time evolution of linear rotors during and after an off-resonant, linearly polarized laser pulse illumination.
 A lot of  work has been done during the last decades employing the impulsive limit for very short pulses, providing analytic expressions in the $\tau_L \rightarrow 0$ limit  with applications to molecular alignment and orientation~\cite{Ortigoso1999,CaiPRL01, cai2001recurring,owschimikow2009state,Mirahmadi_2018,Mirahmadi_2021}, controlling molecular vibrational states~\cite{LemeshkoPRL09, Lemeshko:2010aa}, as well as studying the dynamics of atoms~\cite{Lugovskoy:2015aa}, semiconductor nanostructures~\cite{Matos-Abiague:2006aa}, and low-dimensional electronic systems driven by  pulses~\cite{MOSKALENKO20171}. 
 Our approach goes beyond these efforts by illustrating how deviations occur from the sudden approximation, providing some understanding of the specific conditions that cause these deviations. Our approach can be extended to more complex molecules with higher order polarizability terms and other laser polarization schemes. While the sudden limit for multi-cycle pulses is well-established~\cite{PhysRevA.77.023407, bitter_experimental_2016,bitter_control_2017}, we also investigate the effects of half-cycle pulses, which can generate unipolar fields~\cite{PhysRevA.65.063408,arkhipov2023unipolar,PhysRevA.104.063101,PhysRevA.105.043103}. Using a theoretical method accounting for the full time-evolution operator, we demonstrate that the validity of the sudden limit can be understood in terms of a critical angular momentum threshold $l_\mathrm{crit}$. We propose a new method involving  rescaling of matrix elements, resulting in an effective theory that accounts for deviations from the standard impulsive limit when encountering extended pulse durations.
Our findings hold significant implications for experimentalists working with ultrashort lasers and theorists who employ the sudden limit within their models.
\section{Method}
\subsection{Ultrashort laser pulses}
Here we focus on time-dependence of the full time-evolution operator instead of time-evolving a single initial state with respect to a given laser envelope, as commonly used to describe the dynamics of rotational wavepackets. The advantage is that we do not only learn about the time-evolution of a particular initial state, but also of all possible superpositions. The rigid rotor Hamiltonian can be written as $H_0=B\hat{\mathbf{L}}^2$ with the squared angular momentum operator $\hat{\mathbf{L}}^2$. The potential energy of a polar rotor in an electromagnetic field is given by  
   $V(t) = - \boldsymbol{\mu}\cdot \boldsymbol{\mathcal{E}}(t)$ with (total) dipole moment $\boldsymbol{\mu}$ and laser field amplitude $\boldsymbol{\mathcal{E}}(t)$. Strong fields can give rise to an induced dipole moment $\mu_i = (\mu_0)_i + \frac{1}{2}\sum_{j} \alpha_{ij} \, {\mathcal{E}}_j(t)+ \mathcal{O}[\mathcal{E}^2(t)]$ with the permanent dipole moment of the molecule $\boldsymbol{\mu_0}$ and the  polarizability tensor  $\alpha_{ij}$. The interaction of a linear molecule with a ultrashort, off-resonant linearly polarized laser pulse is given by~\cite{dion_laser-induced_1999,dion2001orienting}
\begin{equation}
  \hat{H}(t) = \hat{H}_0 - \mu_0 \mathcal{E}(t)\cos(\hat{\theta})- \tfrac{1}{4}\mathcal{E}^2(t) \Delta \alpha \cos^2(\hat{\theta}) 
\end{equation}
with angle between field polarization and molecular axis $\theta \in [0,\pi] $, the electric field in the $Z$-direction $\mathcal{E}(t)$ and the difference between parallel and perpendicular polarizability $\Delta \alpha$. \\\\
In the far-field limit the electric field of the laser pulse has to integrate to zero~\cite{rauch2006time,arkhipov2022half,barth2008nonadiabatic}
\begin{equation}
  \int_{-\infty}^{\infty}\mathcal{E}(t)\d t=0.
  \label{eq:int_zero}
\end{equation}
For a laser pulse with many cycles one often assumes that only the part with $\mathcal{E}(t)^2$ is relevant, since the linear term averages out. In that case, one can assume a purely positive Gaussian shape $\mathcal{E}(t)>0$ for the laser field amplitude with kick strength $P_2$, peak position $t_0$ and width $\sigma_t$. In the sudden approximation, the time-evolution propagator (for $t\gg t_0$) takes the simple form 
\begin{equation}
   \hat{U}_\mathrm{sudd, gaussian}  = e^{-\im \hat{H}_0 (t-t_0)/\hbar}e^{+\im P_2 \cos^2(\hat{\theta})}e^{-\im \hat{H}_0 t_0 /\hbar}.
   \label{eq:sudden1}
\end{equation}
Note that the kick strength is dimensionless and can be calculated as~\cite{leibscher2003molecular}
\begin{equation}
    P_2 = -\frac{\Delta \alpha}{4\hbar}\int_{-\infty}^{\infty}\mathcal{E}^2(t)\d t
\end{equation}
Although it is possible to replace pulses with kicks, for few- and half-cycle pulses one has to take into account the full spatial dependence of the laser field. Here, we analyze the half-cycle pulse as an exemplary and experimentally important case, but this analysis can be extended straightforwardly to few-cycle pulses. We consider the following parametrization from Ref.~\cite{barth2008nonadiabatic}: 
\begin{equation}
   \mathcal{E}(t) = 
\begin{cases}
0 & (t \leq 0)\\
\mathcal{E}_1 \cos^2(\omega_L (t-t_p)/2)\sin(\omega_L(t-t_p)) & (0 \leq t < t_p)\\
\mathcal{E}_2 \left(1 - e^{-(t-t_p)/\tau_1} \right) e^{-(t-t_p)/\tau_2} & (t \geq t_p),
\end{cases}
\label{eq:pulse_shape}
\end{equation}
with electric field amplitudes $\mathcal{E}_1, \mathcal{E}_2>0$, the laser frequency $\omega_L$, the pulse duration of the first part of the laser pulse $t_p=\pi/\omega_L$ (in the following referred to as positive pulse duration), the switch-on and switch-off times $\tau_1, \tau_2$. The ratio 
\begin{equation}
  \xi \equiv \mathcal{E}_2/ \mathcal{E}_1  
  \label{eq:xi}
\end{equation}
determines the width of the first peak relatively to the negative tail. 
\begin{figure}[b]
    \centering
    \includegraphics[width=0.5\textwidth]{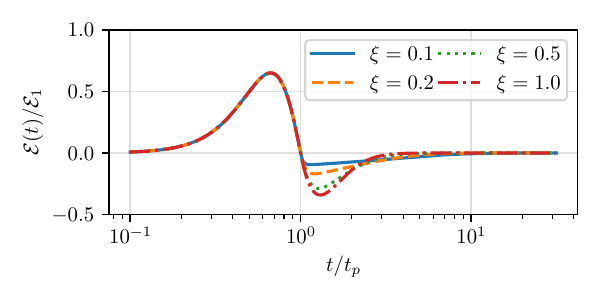}
    \caption{Parametrization of a half-cycle pulse as  given by Eq.~\eqref{eq:pulse_shape} in units of the pulse position $t_p$. $\mathcal{E}_1$ determines the pulse maximum, $\mathcal{E}_2$ the minimum and the ratio $\xi=\mathcal{E}_2/\mathcal{E}_1$ the decay time (see the text). The laser pulse duration $\tau_L$ includes the negative tail up to some degree depending on the field strength. In the Gaussian approximation, the pulse width is approximately given by $\tau_L \approx t_p$.}
    \label{fig:}
\end{figure}
The condition that the electric field is smooth at $t=t_p$ further leads to $\tau_1 = \frac{\mathcal{E}_2}{\omega_L \mathcal{E}_1}=\xi/\omega_L$ and Eq.~\eqref{eq:int_zero} leads to 
\begin{equation}
\begin{aligned}
  \tau_2 &= (2\omega_L^2\tau_1)^{-1} + \sqrt{(2\omega_L \tau_1)^{-2} + (\omega_L)^{-2}} \\ &=  (2\omega_L \xi)^{-1} + \sqrt{(2 \xi)^{-2} + (\omega_L)^{-2}},
\end{aligned}
\end{equation}
see Ref.~\cite{barth2008nonadiabatic}. The decay time is determined by $\tau_2$. The sudden limit for this potential follows as
\begin{equation}
    \hat{U}_\mathrm{sudd, half-cycle}  = e^{-\im \hat{H}_0 (t-t_0)/\hbar}e^{+\im P_1 \cos(\hat{\theta})}e^{-\im \hat{H}_0 t_0 /\hbar}
    \label{eq:sudden2}
\end{equation}
with estimated peak position $t_0$ and kick strength $P_1$. Observe that $t_0$ does not have to match with $t_p$, as the pulse's peak (\ie the pulse position) occurs for $t_0 < t_p$. Furthermore, the duration $t_p$ might not align with the laser duration $\tau_L$ based on the value of $\xi$, since it would disregard the negative tail of the pulse. Still $P_1$ is frequently approximated in the literature as~\cite{dion2001orienting}
\begin{equation}
P_1 \approx \frac{\mu_0}{\hbar}\int_{-\infty}^{t_p}\mathcal{E}(t)\d t,
\label{eq:estimation_P1}
\end{equation}
\ie~by the integral over the positive part of the field amplitude. This is a good approximation when the half-cycle pulse looks similar to a Gaussian pulse, which we demonstrate below. Approximately, the integral over  the positive peak scales as $P_\mathrm{1}\propto \mathcal{E}_1 \cdot t_p$ (the negative tail compensates for exactly this
value). Molecular rotation sets the timescale of the Hamiltonian, thereby justifying the representation of time in units of the rotational revival time $\tau_B= \pi \hbar/B$, denoted as $\tilde{t} = t/\tau_B $. In an effort to render the Hamiltonian dimensionless, we can conveniently incorporate the $\hbar^{-1}$ prefactor of the time evolution into the coupling constants, resulting in the following expression~\footnote{Note that we do not employ the common units of $H/B$, since we are interested in expressing time in units of $\tau_B$, which leads to an additional factor of $\pi$ in front of $\hat{\mathbf{L}}^2$,}
\begin{equation}
  \tilde{H}(\tilde{t})= \pi \hat{\mathbf{L}}^2 - \mathcal{E}(\tilde{t})/\mathcal{E}_\mu \cos(\hat{\theta}) - \mathcal{E}^2(\tilde{t})/\mathcal{E}^2_{\Delta \alpha} \cos^2(\hat{\theta}).
  \label{eq:rescaling_units}
\end{equation}
This includes the constants 
\begin{equation}
  \mathcal{E}_\mu = \frac{B}{\pi \mu}, \quad \mathcal{E}_{\Delta \alpha}=\sqrt{\frac{4 B}{\pi \Delta \alpha}},
  \label{eq:specific_constants}
\end{equation}
which depend on the particular molecule under study (see Section~\ref{sec:wavepacket} for an illustrative example of a time-evolution for the molecule OCS). Moving forward, we will omit the tilde on $t$ and $H$, keeping in mind that all expressions are now unitless. \\\\
In order to study the validity of the sudden approximation, we numerically integrated the differential equation of the time-evolution operator $\hat{U}_\mathrm{full}(t)$,
\begin{equation}
\im \partial_t \hat{U}_\mathrm{full}(t) = \hat{H}(t)\hat{U}_\mathrm{full}(t)  
\end{equation}
for a reasonable cutoff $l < l_\mathrm{max}$ and various parameters~\footnote{For each calculation we increase the cutoff scale until the results we are interested in are converged. This typically depends on the timescale (since high $l$ correspond to high frequency) and the field strength (which determines how many $l$ states are occupied).}. As mentioned in the introduction, each angular momentum eigenstate $|l,m \rangle$ oscillates with the frequency 
\begin{equation}
 \omega_\mathrm{rot}(l)=\pi \cdot l(l+1)/\tau_B
\end{equation}
which provides a natural cutoff scale; the approximation can only succeed for states with $\langle l|\psi \rangle \approx 0$ for $l$ with $\tau_L > \tau_\mathrm{rot}(l)$. The eigenstates $l$ with $\tau_L > \tau_\mathrm{rot}(l)$ oscillate with a frequency equal or higher than the pulse duration and a separation of timescales is not possible. The matrix elements for the potentials are 
    \begin{align}
      \langle l'm'|\cos(\theta)|lm\rangle &= -\delta_{mm'}C^{l'm}_{lm10}C^{l0}_{l'010} \label{eq:exact1}\\
      \langle l'm'|\cos^2(\theta)|lm\rangle &=+\delta_{mm'}\left (\tfrac{2}{3}C^{l'm}_{lm20}C^{l0}_{l'020} + \tfrac{1}{3}\delta_{ll'} \right),
      \label{eq:exact2}
    \end{align}
with the  Clebsch–Gordan coefficients $C_{lml'm'}^{LM}$~\cite{varshalovich1988quantum}. Henceforth, our analysis will concentrate exclusively on linearly polarized laser fields for which different $m$-sectors are independent and we can assume $m=0$. Following the definitions for the sudden limit in Eqs.~\eqref{eq:sudden1} and~\eqref{eq:sudden2}, the effective potential of the full time evolution can be calculated as
\begin{equation}
    \hat{V}_\mathrm{eff}(t) = -\im\log[e^{+\im \hat{H}_0 (t-t_0)}  
\hat{U}_\mathrm{full}(t)\,e^{+\im \hat{H}_0 t_0}]
\label{eq:Veff}
\end{equation}
where one has to use the correct branch cut of the logarithm~\footnote{For values of the effective kick strength smaller than $P \approx \pi$ the logarithm is straightforward to calculate. For larger values one has to resort an algorithm that guarantees a smooth transition of the operator eigenvalues in order to choose the correct branch cut.}.
For times $t \ll t_0$ it converges to a constant, time-independent potential $\hat{V}_\mathrm{eff}\equiv \hat{V}_\mathrm{eff}(t=\infty)$. This is the potential an instantaneous laser pulse at $t_0$ exerts upon the molecule, after the full time evolution. We want to know if the effective matrix elements resemble the ones given in~\eqref{eq:exact1} and~\eqref{eq:exact2}. For perfect agreement the off-diagonal matrix elements
\begin{equation}
   v^{(s)}_l = \langle l\pm s|\hat{V}_\mathrm{eff}|l\rangle \quad \mathrm{with}\,\,  s\in\{1,2\}
\end{equation}
should resemble  $P_\mathrm{s}\cdot \langle l\pm s|\cos^s(\hat{\theta})|l\rangle $ where $P_\mathrm{s}$ depends on the field $\mathcal{E}(t)$. In that case, we can find the strength by $P_\mathrm{s} = v_l^{(s)}/ \langle l\pm s|\cos^s(\hat{\theta})|l\rangle $ which should be the same for all $l$. However, in a realistic case the matrix elements deviate from that obtained in the sudden limit. This implies that the kick strength coefficients
\begin{equation}
  p^{(s)}_l \equiv v_l^{(s)}/\langle l\pm s|\cos^s(\hat{\theta})|l\rangle  
  \label{eq:pl}
\end{equation}
depend on $l$. In many cases, we are only interested in the convergence up to some experimentally relevant $l_\mathrm{av}$. We define the  average of a matrix element $A_l$ as $\bar{A} \equiv \frac{1}{l_\mathrm{av}+1}\sum_{l=0}^{l_\mathrm{av}}A_l$
and estimate the strength $P_{s,\mathrm{eff}}$ and its error by
\begin{equation}
   P_{s,\mathrm{eff}} \equiv \overline{p^{(s)}} , \quad \delta P_{s,\mathrm{eff}} \equiv  \sqrt{\overline{ (p^{(s)})^2} - \overline{p^{(s)} }^2}.
   \label{eq:averages}
\end{equation}
Clearly, if the sudden approximation was exact we would find $\delta P_{s,\mathrm{eff}}=0$. For the case, where the sudden approximation is applicable, this value should be sufficiently small. However, for small kick strengths, this error becomes small as well, therefore, it is necessary to consider the relative error 
\begin{equation}
r_{s}\equiv \delta P_{s,\mathrm{eff}}/P_{s,\mathrm{eff}}.
\label{eq:rel_error} 
\end{equation}
Only the size of $r_{s}$ poses a sufficient criterion whether the sudden limit approximation is valid or not. Until now we have assumed that we are looking at the impulsive limit in the form of Eqs.~\eqref{eq:sudden1} and~\eqref{eq:sudden2}. However, there is a more generic possibility of 
\begin{equation}
 \hat{U}_\mathrm{sudd, generic}  = e^{-\im \hat{H}_0 (t-t_0)}e^{+\im \hat{V}_\mathrm{eff}}e^{-\im \hat{H}_0 t_0}
 \label{eq:sudden3}
\end{equation}
with $\hat{V}_\mathrm{eff}$ as defined in Eq.~\eqref{eq:Veff}. In particular, as we will see later, the numerically estimated effective potentials will often have the same off-diagonal structure as the generating potentials $\hat{V}(t)$. Therefore, it is possible to use \textit{rescaled} matrix-elements $v_l^{(s)}$ that originate from finite time pulses or pulses that are not Gaussian, such as  half-cycle pulse. A rescaled potential will have the form $v^{(s)}_l\rightarrow v^{(s)}_l f^{(s)}_l $ with some function $f^{(s)}_l$ that depends on the laser shape. For Gaussian pulses we can find $f^{(2)}_l$ straightforwardly by
\begin{equation}
   f^{(2)}_l = p^{(2)}_l / P_2.
   \label{eq:fl2}
\end{equation}
with the error factor
\begin{equation}
   \delta_l = 1 - f^{(2)}_l 
   \label{eq:delta}
\end{equation}
\begin{figure*}
    \centering
    \includegraphics[width=1.0\textwidth]{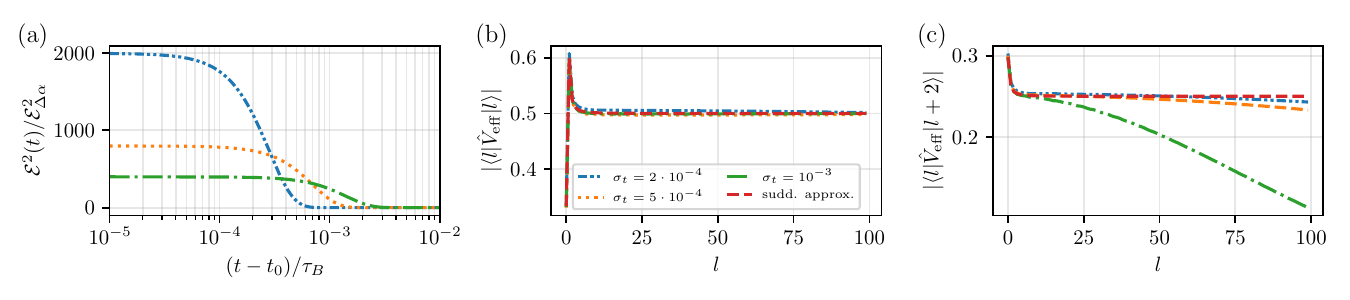}
    \caption{Results for a Gaussian pulse with the $\cos^2(\hat{\theta})$ term and $\mu_0=0$. The full time evolution was integrated numerically and $\hat{V}_\mathrm{eff}$ was calculated using Eq.~\eqref{eq:Veff}. In (a) the field strength squared after the peak, in (b) the diagonal matrix element and in (c) the second off-diagonal matrix element. Other matrix elements are close to zero. We observe that the diagonal matrix elements (b) coincide perfectly with the sudden limit (red dashed line), but the second off-diagonal matrix elements (c) show large deviations. For increasing $\sigma_t$, the deviations set in for lower $l$. It becomes clear that there only for $l<l_\mathrm{crit}$ with some $l_\mathrm{crit}(P,\sigma_t)$ the sudden limit with $V_\mathrm{eff}=P \cos^2(\hat{\theta})$ is a valid approximation.}
    \label{fig:gaussian1}
\end{figure*}
\begin{figure*}
    \centering
    \includegraphics[width=1.0\textwidth]{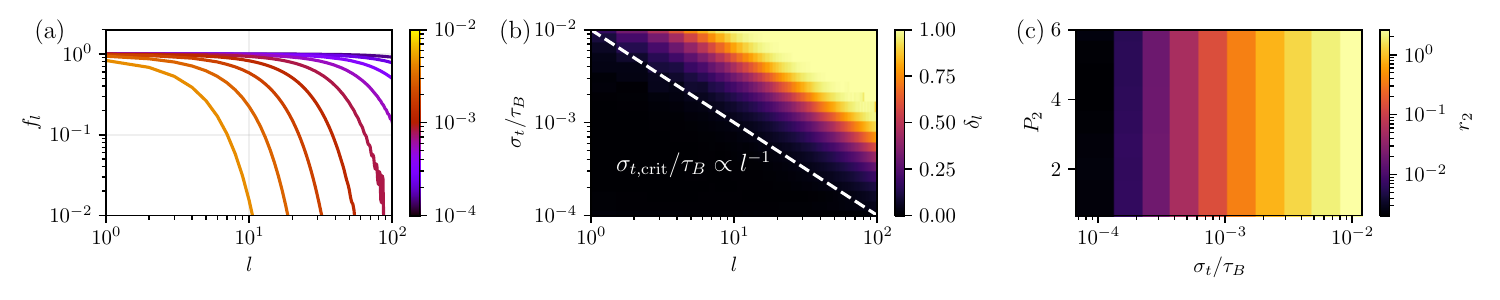}
    \caption{(a) The scaling factor $f_l$, Eq.~\eqref{eq:fl2}, for different pulse widths $\sigma_t$ as indicated by the color; (b) the deviation $\delta_l$, Eq.~\eqref{eq:delta}, for a Gaussian pulse with $P_2=1$. In the region with $f_l \approx 1$ or low error $\delta_l$ (black) the sudden limit is a good approximation, \ie for $l \ll l_\mathrm{crit}$. Approximately above the dashed white line (\ie the non-black region) with $\sigma_t \propto l^{-1}$, there are large deviations from a delta $\cos^2(\theta)$ potential, see Fig.~\ref{fig:gaussian1} for examples. (c) The relative error $r_2$, see Eq.~\eqref{eq:averages}. This figure demonstrates that the goodness of the approximation is independent of $P_2$, \ie the integral of the pulse: it is only sensitive to the width $\sigma_t$.}%
    \label{fig:gaussian2}
\end{figure*}

that gives a good indication how much rescaling is necessary. For half-cycle pulses such a simple expression is not possible, since one has to infer additionally the effective strength $P_1$.
We introduce the usual interaction picture of a Hermitian operator $\hat{A}$ by
\begin{equation}
\hat{A}_I(t) = e^{+i \hat{H}_0 t} \hat{A} e^{-i \hat{H}_0 t}
\label{eq:interactionpicture}
\end{equation}
and the time-evolution operator (with $t_0=0$) with  $U_I(t)= e^{+i \hat{H}_0 t}\hat{U}(t)$.
The Schrödinger equation then reads 
\begin{equation}
   i \partial_t  \hat{U}_I(t) = \hat{V}_I(t)\hat{U}_I(t).
   \label{eq:schroedinger}
\end{equation}
In the following we resort to numerical integration of ~\eqref{eq:schroedinger} and use~\eqref{eq:Veff} to calculate the effective potential directly.
\section{Results}
In this section, we scrutinize the application of the sudden approximation to multi-cycle pulses. Following this, we turn our attention to the analysis of half-cycle pulses. Notwithstanding the disparities in laser frequencies between these pulse types -- optical frequencies for multi-cycle pulses and terahertz frequencies for half-cycle pulses -- similar impacts are discerned in their interaction with rotors.
\subsection{Gaussian pulses}
We model multi-cycle pulses using Gaussian functions $\mathcal{E}^2(t)/\mathcal{E}^2_{\Delta \alpha} = e^{-(t-t_0)^2/2\sigma_t^2}/(\sigma_t \sqrt{2\pi})$ with the squared field strength $\mathcal{E}^2(t)$, $\mu_0=0$, and $P_2=1$, which we will denote \textit{Gaussian pulses} in what follows. Hence, the pulse width of the laser can be directly inferred from $\tau_L \approx \sigma_t$, depending on the definition of $\tau_L$.\\\\
Figure~\ref{fig:gaussian1} provides an illustration of the results calculated for a range of $\sigma_t$. As one would intuitively expect we observe that as the ratio $\sigma_t/\tau_B$ becomes increasingly small, the approximation aligns more closely with the sudden limit. However, with an increase in the value of $\sigma_t$, the effective potential begins to display noticeable deviations from the sudden limit. This divergence is prominently displayed in the off-diagonal matrix elements.\\\\
A detailed look at these matrix elements reveals a significant decrease for larger values of $l$. This contrasts with the matrix elements of the pure sudden pulse, which remains constant. One of the primary features of the perfect delta kick is its ability to transfer angular momentum even for states with high $l$ values. However, this feature is absent in the case of pulses of finite width. Here, the transfer of angular momentum may cease altogether for large $l$. This can occur when the rotational periods $\tau_\mathrm{rot}(l)$, are comparable or smaller than the laser pulse duration $\tau_L$. We assume that such parity leads to destructive interference, inhibiting the laser's capacity to transfer energy to the molecule coherently.\\\\
The phenomenon is more clearly depicted in Figure~\ref{fig:gaussian2}, where the scaling factor, $f_l$, and its error, $\delta_l \equiv 1 - f_l$, are showcased for different values of $\sigma_t$. When the values of $f_l$ or $\delta_l$ are equal to 1 or 0 respectively, it indicates an agreement with a delta kick. However, if $\delta_l$ diverges from 0, it signals a deviation from a delta kick. As per our findings, the sudden limit holds true until a certain critical value, $l_\mathrm{crit} \propto \sigma_t^{-1}$. Once this point is surpassed, the sudden limit no longer applies, leading to decay in matrix elements and rapid growth in deviations.\\\\
Henceforth, the time-evolution of a wavepacket that is driven by a Gaussian-shaped pulse can be captured by the sudden approximation when the wavepacket has only occupations for $l<l_\mathrm{crit}(\sigma_t)$. In that case, the sudden approximation is valid and it is not necessary to integrate the Schrödinger equation fully. Another possibility is to rescale the effective potential to
\begin{equation}
   \langle l'm'|\hat{V}_\mathrm{rescaled}|lm\rangle = \delta_{mm'}\left( f^{(2)}_l\tfrac{2}{3}C^{l'm}_{lm20}C^{l0}_{l'020} + \tfrac{1}{3}\delta_{ll'} \right)
\end{equation}
with the rescaling function $f_l$. This way we can capture the deviations that arise due to the non-zero pulse width. However, this rescaling is not possible in all cases, as we will demonstrate in Section~\ref{sec:wavepacket}. These findings are expected to be useful in understanding the behavior of half-cycle pulses, which we will be exploring in our subsequent analysis.
\subsection{Half-cycle pulses}
\begin{figure*}
    \centering
    \includegraphics[width=0.8\textwidth]{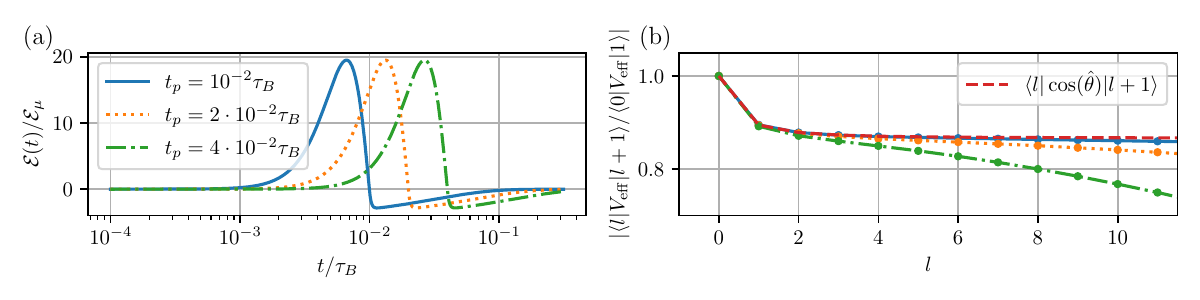}
    \caption{Behavior of a finite-width half-cycle pulse, Eq.~\eqref{eq:pulse_shape}, for different positive pulse durations $t_p = \pi/\omega_L$, where $t_p$ also controls the peak width and hence also the strength $P_1$. (a) The field strength $\mathcal{E}(t)/\mathcal{E}_\mu$ for $\mathcal{E}_1/\mathcal{E}_\mu=30$ and $\mathcal{E}_2/\mathcal{E}_\mu=0.03$. (b)~The first off-diagonal matrix element of the effective potential defined in Eq.~\eqref{eq:Veff}. Note that we divide by $\langle 0 | V_\mathrm{eff}|1\rangle $ to bring the potentials on top of each other since each potential corresponds to a different $P_1$. Similar to the multicycle pulses of Fig.~\ref{fig:gaussian1} we observe that for small $t_p$ the matrix elements coincide perfectly with the sudden limit (red dashed line).}%
    \label{fig:halfcycle1}
\end{figure*}
\begin{figure*}
    \centering
    \includegraphics[width=1.0\textwidth]{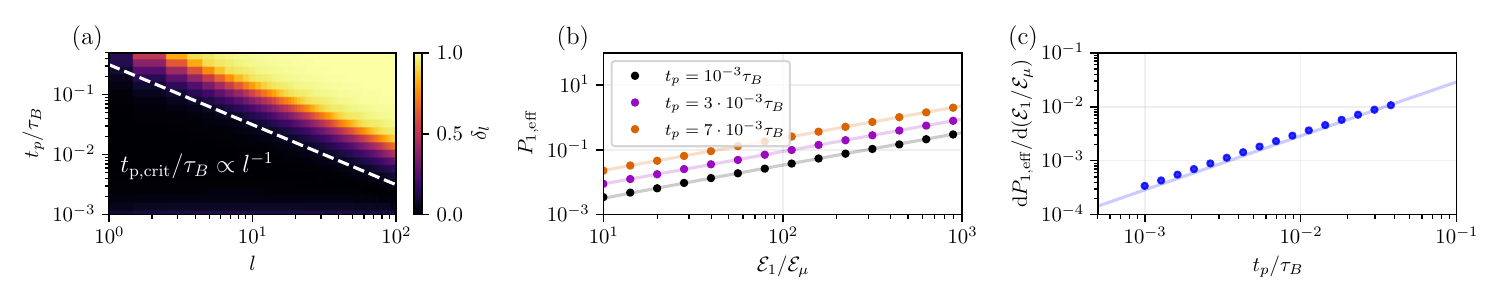}
    \caption{Outcomes for half-cycle pulses in the \textit{Gaussian} limit. (a)~The deviation error, denoted as $\delta_l = |1 - p_{l}^{(2)}/p^{(2)}_{l=0}|$, for a half-cycle pulse where $\xi=10^{-3}$ (corresponding to the Gaussian limit) and $l_\mathrm{max}=250$. Similar to the Gaussian pulse (refer to Fig.~\ref{fig:gaussian2}), a power-law dependence on the critical pulse width is observed (indicated by the dashed white line, estimated visually). (b)~The effective kick strength, $P_\mathrm{1,eff}$ evaluated up to $l_\mathrm{av}=50$, as a function of field strength $\mathcal{E}_1/\mathcal{E}_\mu$ for three distinct $t_p$. (c)~Confirms the anticipated linear relationship, $P_\mathrm{1,eff} \propto t_p \cdot \mathcal{E}_1 $, with $\frac{\d^2 P_\mathrm{1,eff}}{\d \mathcal{E}_1 \d t_p} \approx 0.287$.} \label{fig:halfcycle2}
\end{figure*}
\begin{figure*}
    \centering
    \includegraphics[width=1.0\textwidth]{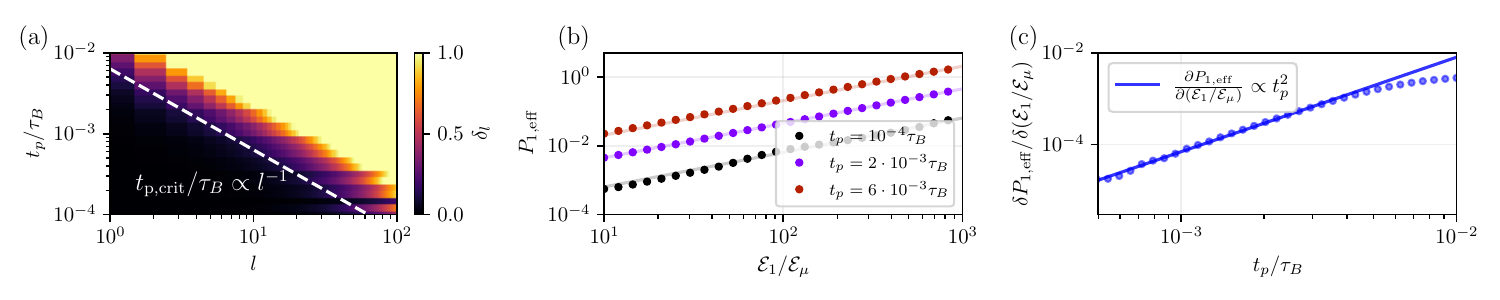}
    \caption{Results for half-cycle pulses for the \textit{oscillating} limit with $\xi=\mathcal{E}_2/\mathcal{E}_1=1$ with the same parameters as in Fig.~\ref{fig:halfcycle2}, \ie $\delta_l = |1 - p_{l}^{(2)}/p^{(2)}_{l=0}|$ with $l_\mathrm{max}=500$. 
    Again, a power-law dependence on the critical pulse width is observed (indicated by the dashed white line, estimated visually). However, unlike for Gaussian pulses the break-down of the sudden limit occurs for smaller $l_\mathrm{crit}$. 
    In (b) the effective kick strength, $P_\mathrm{1,eff}$ evaluated up to $l_\mathrm{av}=50$, as a function of field strength $\mathcal{E}_1/\mathcal{E}_\mu$ for three distinct $t_p$. 
    In (c) we demonstrate that the slopes of (b) are related to $t_p$ by $\partial P_\mathrm{1,eff}/\partial (\mathcal{E}_1/\mathcal{E}_\mu)\propto t_p^2$ as long as $t_p$ is not large enough (note that deviations for $t_p \sim 10^{-2}\tau_B$). This is a new result and originates from the non-Gaussian pulse shape, which does not allow for the simple estimation of $P_1$, Eq.~\eqref{eq:estimation_P1}.}
    \label{fig:halfcycle3}
\end{figure*}
\begin{figure*}
    \centering
    \includegraphics[width=1.0\textwidth]{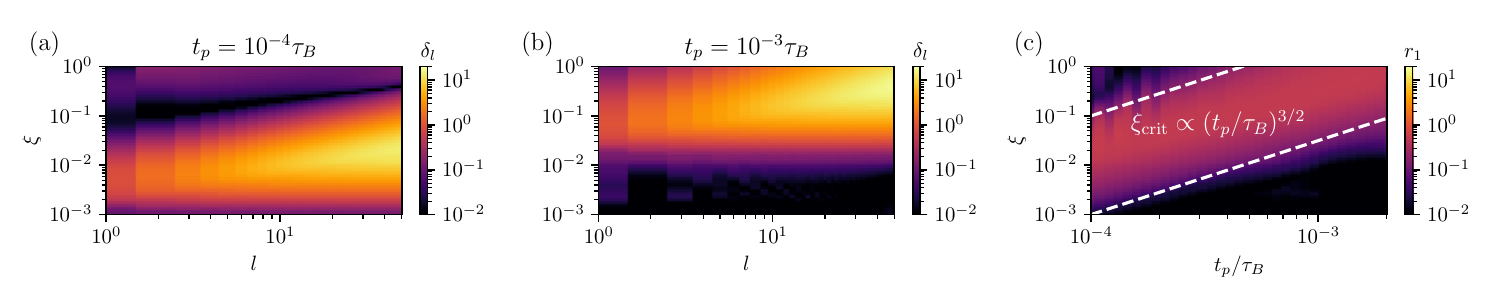}
    \caption{The behavior for an arbitrary value of $\xi$ with $\mathcal{E}_1/\mathcal{E}_\mu=10^3$ proves more intricate than the scenarios we have previously examined. In (a) and (b) the relationship between angular momentum $l$ and $\xi$ for two positive pulse widths $t_p$ that incorporate a large cutoff at $l_\mathrm{max}=500$. As $\xi$ approaches either extreme of 1 or 0 for small $t_p$, we recover the behavior $\delta_l \rightarrow 0$ from previous figures. However, within these extremes, the error scaling is heavily influenced by the positive pulse width $t_p$. This makes sense as the specific timescale exerts a significant impact on how the rotational modes interact with the field. In (c) we can deduce the dependency of the relative error $r_1$, as defined in equation~\eqref{eq:rel_error}, on both $t_p$ and $\xi$ (where we averaged up to an value of $l_\mathrm{av}=20$). We observe that when $\xi$ approaches 1, the condition of $t_p \rightarrow 0$ becomes critical. Interestingly, an increase in $t_p$ values not only allows, but also appears to encourage, higher $\xi$ values when transitioning from the $\xi \approx 0$ limit. We have visually approximated this relationship as $\xi_\mathrm{crit} \propto (t_p/\tau_B)^{3/2}$, although the actual dependence can be more complex. Nevertheless, it's noteworthy that increasing the ratio $t_p/\tau_B$ permits the use of a greater $\xi$ value for fixed $\mathcal{E}_1$.}
    \label{fig:xi_tp}
\end{figure*}
In this section, we shift our focus to half-cycle pulses. For simplicity we focus only on the dominant term, the permanent dipole term with finite $\mu_0>0$. For many linear molecules this is a good approximation since the specific constants~\eqref{eq:specific_constants} satisfy $\mathcal{E}_{\mu} \ll \mathcal{E}_{\Delta \alpha}$. As previously mentioned, in the case of half-cycle pulses, there is a positive peak followed by a potentially long negative tail. The $\langle l|\cos(\hat{\theta})|l' \rangle$ matrix element is only non-zero for $l=l'\pm 1$. In many cases, this is also true for $\hat{V}_\mathrm{eff}$. Specifically, in the limit where the ratio $\xi\rightarrow 0$ from~\eqref{eq:xi}, which we will refer to as the \textit{Gaussian limit}, the behavior converges to the Gaussian pulse discussed earlier, since the depth of the negative tail is minimal and it requires an infinite amount of time to satisfy Equation~\eqref{eq:int_zero}. In Fig.~\ref{fig:halfcycle1}, we illustrate the shape of the potential for very small values of $\xi$. As expected, the effective potential matrix elements diverge from the $\cos(\theta)$ potential for increasing $t_p$, exhibiting similar behavior to that of Gaussian pulses (cf.~Fig.~\ref{fig:gaussian2}). For half-cycle pulses, the positive pulse duration $t_p$ plays a role analogous to the width $\sigma_t$ for Gaussian pulses.\\\\
In Fig.~\ref{fig:halfcycle2} we find that the critical positive pulse width scales as $t_\mathrm{p,crit}/\tau_B \propto l^{-1}$, which is similar to the critical pulse width for Gaussian pulses in Fig.~\ref{fig:gaussian2}. The primary difference arises from the fact that $t_p=\pi/\omega_L$ (only for half-cycle pulses) with the laser frequency $\omega_L$, corresponding to exactly half a cycle, while the variable $\sigma_t$ of the Gaussian pulses corresponds to the width of one standard deviation, or approximately $68\%$ of the nominal pulse area. We note that in the \textit{Gaussian limit}, we do not observe a dependency of the relative error on the kick strength $P_\mathrm{1,eff}$. However, when leaving the Gaussian limit, \ie when $\xi$ is not small, it plays an important role how the potential deviates from the impulsive limit. \\\\
Now we look at the opposite limit $\xi= 1$, which we denote the \textit{oscillating limit}, since the negative tail can not be integrated out, like we did effectively for the Gaussian limit. Also in that limit we find that it is possible to approximate the full time-evolution with the impulsive limit, see Fig.~\ref{fig:halfcycle3}. The main difference is that for a given $t_p$, the sudden approximation breaks down for smaller $l$, which implies that one has to choose smaller widths $t_p/\tau_B$ than in the Gaussian limit to achieve the same accuracy. Further, it is important to note that unlike the Gaussian case the diagonal elements are not vanishing completely. While we confirm the relationship $P_\mathrm{1,eff}\propto \mathcal{E}_1/\mathcal{E}_\mu$, the dependency on $t_p$ is more complicated than in the $\xi\rightarrow 0$ case and we find $\partial P_\mathrm{1,eff}/\partial (\mathcal{E}_1/\mathcal{E}_\mu)\propto t_p^2$, displaying a strong deviation from the generally accepted result~\eqref{eq:estimation_P1}.\\\\ 
Finally, we turn our focus to the case involving arbitrary $\xi$. Our compiled results are presented in Fig.~\ref{fig:xi_tp}. This consolidates our previous analyses for the two limiting scenarios: $\xi \rightarrow 0$ and $\xi=1$. Additionally, it provides an understanding of how the Gaussian and oscillating limits respectively cease to hold for mid-range values of $\xi$, where the error $\delta_l$ grows large already for small $l$. Evidently, in the scenario of $\xi \rightarrow 1$, a small $t_p/\tau_B$ ratio is necessary to maintain the sudden approximation, as has been demonstrated in Fig.~\ref{fig:halfcycle3}. Contrarily, we discover that in the opposing extreme where $\xi \approx 0$, a larger $t_p$ proves beneficial, at least for the relative error. 
\section{Wavepacket time-evolution of OCS}
\label{sec:wavepacket}
\begin{figure*}
\centering
\includegraphics[width=1.0\textwidth]{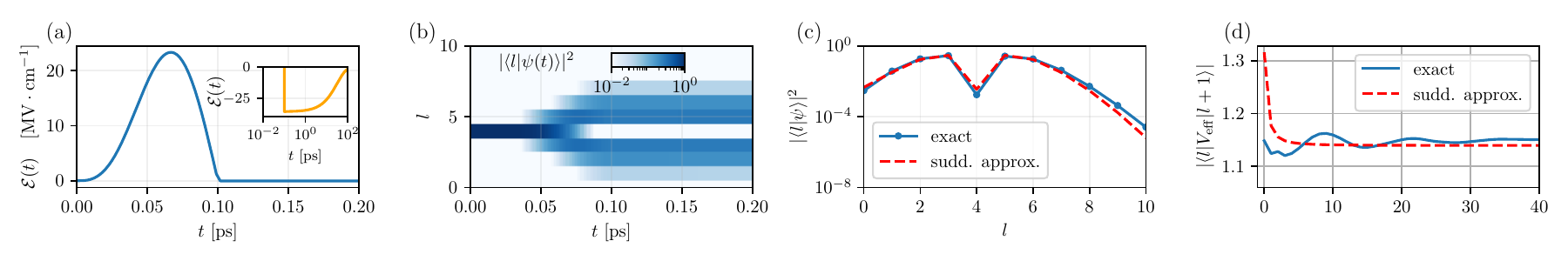}
\caption{
Numerical simulation of wave-packet time-evolution for an OCS molecule. The parameters characterizing the molecule~\cite{tanaka1984co2} are $\tau_B \approx 80$~ps, $\Delta \alpha \approx 4.67$ \r{A}$^3$, and $\mu \approx 0.66$~Debye. Field constants specific to OCS, Eq.~\eqref{eq:specific_constants}, are used.
(a) The field profile of the  half-cycle laser pulse in the Gaussian regime, $\mathcal{E}_1=6000, \xi=10^{-3}$, resulting in a peak intensity $\mathcal{E}_\mathrm{max}\approx 23$ MV/cm. The inset show the long-time behavior of the pulse.
(b) The absolute value of the wavepacket components for a wavepacket  initialized with $l=4$. (c) The converged wavepacket long after the pulse.
(d) The effective matrix elements from the full time-evolution and the sudden approximation for $P_\mathrm{1,eff} \approx 2.28$, as determined by Eq.~\eqref{eq:averages}. The relative error for this potential, calculated using Eq.~\eqref{eq:rel_error}, is a modest $r_1\approx 2 \%$, indicating that the sudden approximation is effective in this context.}
\label{fig:time_evolution1}
\end{figure*}

\begin{figure*}
\centering
\includegraphics[width=1.0\textwidth]{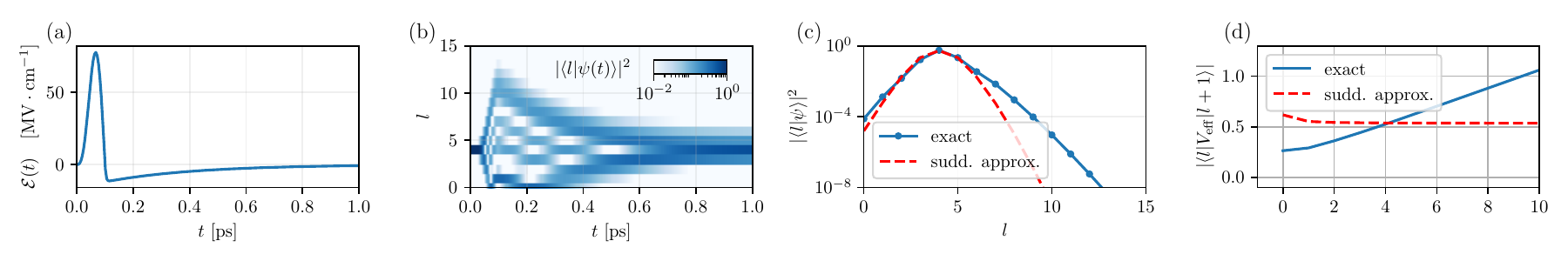}
\caption{
Same as in Fig.~\ref{fig:time_evolution1}, but in the intermediate regime with $\xi=0.1$. (b) Shows that the negative tail now lowers the occupations at high $l$ values and (c) demonstrates the deviation from the sudden approximation. 
(d) The effective matrix elements for $P_\mathrm{1,eff}\approx 1.07$, calculated by Eq.~\eqref{eq:averages} for matrix elements up to $l_\mathrm{av}=10$. The substantial relative error of $r_1\approx 43 \%$ indicates the inadequacy of the sudden approximation in this case.}
\label{fig:time_evolution2}
\end{figure*}
\begin{figure*}
\centering
\includegraphics[width=1.0\textwidth]{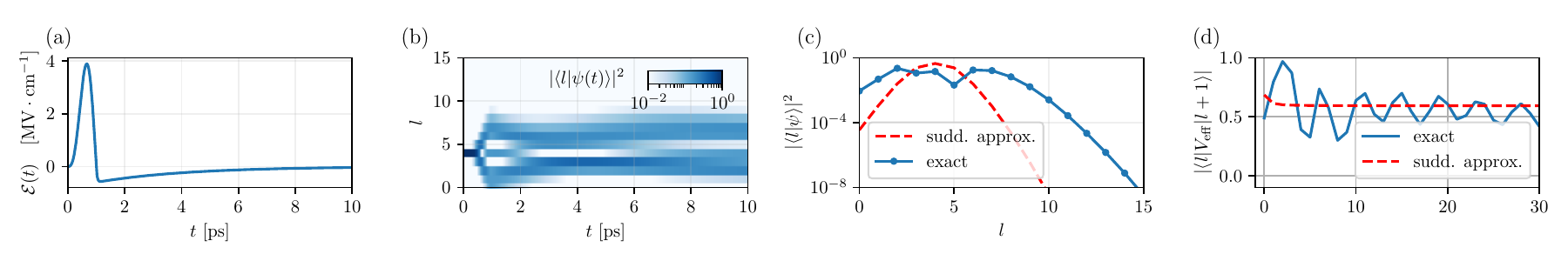}
\caption{
Same as in Fig.~\ref{fig:time_evolution1}, but with a significantly extended pulse duration of $t_p=1$~ps.
(d) The effective matrix elements for $P_\mathrm{1,eff} \approx 1.2$ as calculated by Eq.~\eqref{eq:averages} for matrix elements up to $l_\mathrm{av}=10$. High relative error of $41 \%$ underscores the poor agreement with the sudden approximation. This discrepancy is attributed to oscillations in the effective potential induced by the pulse width, which is comparable to the rotational periods $\tau_\mathrm{rot}(l)$ of some non-zero angular momentum states.}
\label{fig:time_evolution3}
\end{figure*}
\begin{figure*}
\centering
\includegraphics[width=1.0\textwidth]{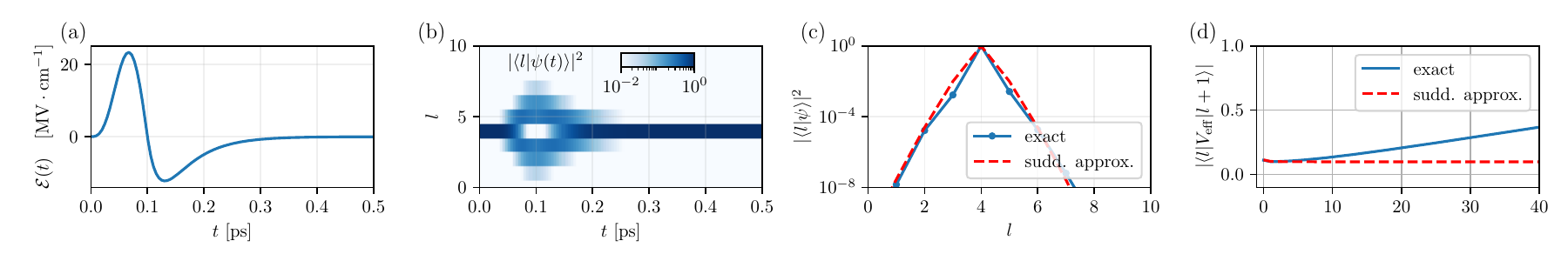}
\caption{
Same as in Fig.~\ref{fig:time_evolution1}, but under oscillating conditions, $\xi=1$.
(b)  highlights how the sharp the negative slope of the pulse counteracts the positive peak, leading to an almost complete negation of previously occupied angular momentum states. Despite this behavior, the agreement with the sudden approximation is very high, as seen in (c) with the long-time result of the wavepacket.
(d) The effective matrix elements for $P_\mathrm{1,eff} \approx 0.2$, as calculated by Eq.~\eqref{eq:averages} for matrix elements up to $l=5$ with a relative error of $3 \%$. }
\label{fig:time_evolution4}
\end{figure*}
We executed a series of numerical simulations, aiming to examine the dynamics of an OCS molecule's wave-packet under illumination of different half-cycle pulses. In Figs.~\ref{fig:time_evolution1},~\ref{fig:time_evolution2},~\ref{fig:time_evolution3}, and~\ref{fig:time_evolution4}, we present the results using $\tau_B \approx 80$~ps, $\Delta \alpha \approx 4.67$ \r{A}$^3$, and $\mu \approx 0.66$ Debye~\cite{tanaka1984co2}. By using rescaled units~\eqref{eq:rescaling_units}, we obtain the
specific field constants $\mathcal{E}_\mu \approx 6$~kV/cm and $\mathcal{E}_\mathrm{\Delta \alpha} \approx 1$ MV/cm. Since $\mathcal{E}_\mu \ll \mathcal{E}_{\Delta \alpha}$, we neglect the influence of the $\Delta \alpha$ term in what follows. In a study by Fleischer et al.~\cite{PhysRevLett.107.163603}, they reported the use of half-cycle pulses with an average field strength of approximately 22 kV/cm up to 1 MV/cm when applied to OCS molecules, which is the regime we are examining here. Note that as can be inferred from Fig.~\ref{fig:halfcycle2}, the relative field strength $\mathcal{E}/\mathcal{E}_\mathrm{\Delta \alpha}$ should be on the order of $100- 1000$ in order to see a visible effect on the molecule.\\\\
The time-dependent wave-packet evolution of a molecule (with $m=0$) is controlled by
\begin{equation}
\partial_t C_l(t) = -i\sum_{l'=0} \langle l'|\hat{V}_I(t) | l \rangle \, C_{l'}(t),
\end{equation}
with the potential in the interaction picture defined in~\eqref{eq:interactionpicture}, and the solution for the wavefunction 
\begin{equation}
\langle l | \psi(t) \rangle = C_l(t)e^{-i \pi l(l+1) t}
\end{equation}
in units of rotational time $\tau_B$.\\\\
In Fig.~\ref{fig:time_evolution1}, the molecule is exposed to a half-cycle pulse in the Gaussian regime, with $\xi=10^{-3}$, whose profile is shown  in Fig.~\ref{fig:time_evolution1}(a). The pulse has a width $t_p$, significantly shorter than the molecule's rotational period. The wavepacket in the initial condition is in a pure $l=4$ angular momentum state, \ie $\langle l|\psi \rangle = \delta_{l,4}$. During the pulse illumination, the pulse performs akin to a Gaussian pulse, with both lower and higher angular momentum states being occupied. Post-illumination, a decrease in the occupation probability for state $l=4$ is evident, possibly due to destructive interference. We use the sudden approximation, defined by~\eqref{eq:averages}, to estimate the effective kick strength of an instantaneous delta pulse. This approximation mirrors the final state of the wavepacket with high precision, demonstrating a fidelity of $97\%$, and it accurately predicts the dip in the $l=4$ state. The sudden approximation's agreement with the full time-evolution is further confirmed by the effective matrix elements~\eqref{eq:Veff}, see Fig.~\ref{fig:time_evolution1}(d). The small relative error~\eqref{eq:rel_error} of $r_1\approx 2 \%$ underscores the appropriateness of the sudden approximation in this context.\\\\
Figures~\ref{fig:time_evolution2},~\ref{fig:time_evolution3}, and~\ref{fig:time_evolution4} were created similarly to Fig.~\ref{fig:time_evolution1}, albeit with varied pulse parameters and widths. In Fig.~\ref{fig:time_evolution2}, the pulse is set in the intermediate regime $\xi=0.1$. The sudden approximation proves challenging to apply in this scenario, as evident in the evolution of the representative wavepacket. The matrix elements of the effective potential begin to diverge for large $l$, failing to plateau like in the case of the sudden approximation. Consequently, finding the correct kick strength that could reproduce the full time-evolution results is problematic. Therefore, we advise against using the sudden approximation in such a scenario due to the significant deviations.\\\\
In Fig.~\ref{fig:time_evolution3}, we adjust the pulse width to a longer duration ($t_p=1$~ps), while staying within the same intermediate regime ($\xi=0.1$). This modification leads to noticeable oscillations (see Fig.~\ref{fig:time_evolution3}(d)) in the matrix elements of the effective potential, resulting from the compatibility of the pulse width with the rotational periods of certain angular momentum states. Evidently, in this regime, the laser's timescale overlaps with the molecule's rotational oscillations, causing interference. This interference hinders the application of the sudden approximation, corroborated by a poor agreement between the wavefunctions of the sudden approximation and the full time evolution (as low as $20\%$). An intriguing observation is the absence of a depopulation in high angular momentum states, likely attributable to the longer duration of the negative peak.\\\\
In the final scenario, as illustrated in Fig.~\ref{fig:time_evolution4}, we look into the oscillating limit by setting $\xi=1$. We observe that the pulse's negative slope almost negates the positive peak, leading to a markedly reduced effective kick strength. Nevertheless, the agreement with the sudden approximation in this regime is remarkably high, presenting a fidelity of $98 \%$, reinforcing our previous analysis of Fig.~\ref{fig:halfcycle3}.\\\\
\section{Conclusions}
In summary, we have assessed the validity of the impulsive limit by examining the full time-evolution operator using a method that solves the time-dependent Schrödinger equation at the operator level. Our findings demonstrate that both Gaussian pulses and half-cycle pulses can be accurately described by the sudden limit, provided that the angular momentum is below the critical threshold $l_\mathrm{crit}$, the pulse width $\sigma_t$ or $t_p$ is significantly smaller than the rotational period $\tau_B$, and for half-cycle pulse the pulse is either in the Gaussian limit ($\xi\rightarrow 0$) or the oscillating limit ($\xi = 1$).\\
\\
This can be used to obtain experimental estimates to reliably realize delta kicks for the cases where   the laser parameters fall within the sudden limit regime and the molecule's angular momentum is not excessively large. Under these constraints, it becomes impossible to differentiate between a delta pulse and a finite-width pulse when examining the matrix elements in the long time limit. However, outside this regime, we observe substantial deviations that can be attributed to the to the time evolution within the pulse width.
\\\\
 Our approach, based on the effective potential \eqref{eq:Veff} is similar to approaches based on the Magnus expansion~\cite{magnus1954exponential} and to the technique presented in  Ref~\cite{MOSKALENKO20171}, which focuses on the expansion of the time-evolution operator in terms of $\tau_L$, and deriving related expressions (see e.g.\ Eq.~(16) of Ref~\cite{MOSKALENKO20171}). However, our method provides new insights for rotational states and the off-diagonal matrix elements of specific molecule-laser interactions. By analyzing the deviation from the sudden approximation, we explicitly show how for increasing pulse widths and laser strengths the behavior deviates from the first-order Magnus expansion.
\\\\
This research serves a dual purpose: elucidating the validity boundaries of the impulsive limit and pinpointing specific circumstances under which deviations from the approximation manifest. Further studies may explore a broader range of pulse shapes, such as, e.g., few-cycle pulses. Moreover, the investigation could extend to quantum numbers other than angular momentum $l$, more intricate polarization schemes, or more complex molecules. Our findings could be applicable to other applications involving THz pulses, such as their interaction with electrons. The enhanced control over molecular dynamics provided by our research might be valuable in fields like ultrafast spectroscopy, laser-induced chemistry, and material processing, where precision is vital for realizing targeted results. The novel viewpoints and methodologies proposed in this study could also inspire further research and innovation in molecular rotational dynamics and related fields.
\\\\
We thank Bretislav Friedrich, Marjan Mirahmadi, and Burkhard Schmidt for insightful discussions.
M.L.~acknowledges support by the European Research Council (ERC) Starting Grant No.~801770 (ANGULON). 
\bibliography{main}
\end{document}